\documentstyle{ioplppt}                
\begin{document}
\newcommand{\ox}{$^{16}$O~}
\newcommand{\ipb}{$^{208}$Pb~}
\title{Evidence of Double Phonon Excitations in $^{16}$O~+~$^{208}$Pb Reaction}
\author{M~Dasgupta\dag, K~Hagino\ddag, C~R~Morton\dag\ftnote{3}{Present
 address - Department of Physics, SUNY, Stony Brook, NY 11794, USA.}, 
D~J~Hinde\dag, J~R~Leigh\dag, N~Takigawa\ddag, H~Timmers\dag and 
J~O~Newton\dag}

\address{\dag\ Department of Nuclear Physics, Research School of
Physical
Sciences and Engineering,  \\
 Australian National University, Canberra, ACT 0200, Australia}
\address{\ddag\ Department of Physics, Tohoku University, Sendai 980--77, 
Japan}

\begin{abstract}

The fusion cross-sections for $^{16}$O~+~$^{208}$Pb, 
measured to high precision, enable the extraction of the distribution
of fusion barriers. This  shows a structure markedly
different from the single--barrier which might be {\it expected} for fusion
of two doubly--closed shell nuclei. The results of exact coupled
channel calculations performed to understand the observations are
presented. These calculations indicate that coupling to a double octupole 
phonon 
excited state in $^{208}$Pb is necessary to explain the experimental 
barrier distributions.

\end{abstract}


\section{Introduction}
The concept~\cite{ro91} that a representation of the
 distribution of barriers~\cite{da83}, encountered by two 
colliding nuclei can experimentally be determined~\cite{xi91}
 from precisely 
measured fusion excitation functions has led to a renewed 
interest~\cite{mo94,st95,le95,bi96} in 
heavy--ion fusion studies near the Coulomb barrier. These experiments 
have shown that the dominant factors, i.e., the nature of couplings, 
affecting the fusion probability can 
be very clearly seen in a barrier distribution representation. 
With the advent of this new tool and the new measurements, it is interesting 
to re--visit the 
problem of fusion of two doubly closed shell nuclei, 
the \ox + \ipb system; detailed analysis~\cite{th89} performed 
previously could not reproduce 
the fusion cross-section and the mean square angular 
momentum by coupled channel calculations. Apart from the interest in the 
reaction process 
itself, another aspect is to identify the state(s) in \ipb which 
couple strongly and thus contribute to the shape of the barrier distribution 
in this reaction. With this knowledge, \ipb can then be used as a probe in 
understanding the barrier distributions in  reactions with other nuclei; 
this is an advantage due to the large $Z$ of $^{208}$Pb, leading to large
 coupling 
strength and hence well {\it resolved} barrier distributions. 
Further, understanding the reaction mechanism may also have implications 
in the reactions to produce  super--heavy 
elements, many of which have \ipb as one of the reaction partners.

\section{Experimental procedure}

The experiments were performed using pulsed \ox beams
from the 14UD Pelletron
accelerator at the Australian National University.
Isotopically enriched targets of lead in the form of $^{208}$PbS of 
thicknesses $\sim$23$\mu gm/cm^2$ and  $\sim$350$\mu gm/cm^2$ evaporated
onto a
$\sim$10$\mu gm/cm^2$ carbon backing were used for the fission--fragment
and evaporation residue measurements respectively.
The fission fragments were detected in two large--area position sensitive 
multiwire proportional counters in the forward and backward hemispheres.
Fission events were identified 
by their energy loss, time--of--flight information with respect to the pulsed 
beam. The angular distributions of the fission--fragments, obtained from the 
position information, were used to calculate the total fission cross-sections 
as detailed in Ref.~\cite{mo95}. 
Two silicon surface barrier detectors located at $\pm$22$^\circ$
were used to detect elastically  scattered events for normalization purposes. 

For the measurements of evaporation residues an aluminium catcher foil
was placed behind the target in order to stop the  
recoiling  products. The evaporation residue cross-section was 
measured by detecting the 
$\alpha$--activity from the decay ER and their daughters, using an annular 
silicon surface barrier detector as detailed in Ref.~\cite{mo95}. 
The total fusion cross-section was obtained by summing the fission-- and 
evaporation residue cross--sections.

\section{Results}

The results of the measurements are presented in two
forms: (i) total fusion excitation functions
and (ii) the function $d^2(E \sigma)/dE^2$, where E is the energy and 
$\sigma$ is the fusion cross-section. It has
been shown~\cite{ro91} that the quantity $d^2(E \sigma)/dE^2$ gives a 
representation of the distribution of barriers, and in the following 
discussions will be referred to as the {\it barrier 
distribution}. This quantity has been extracted from the fusion data using 
the point difference formula~\cite{xi91} with a step length of 1.86~MeV
 in centre--of--mass frame.
The theoretical excitation functions have been
treated in exactly the same way as the experimental data to obtain the
theoretically expected barrier distributions. Calculations
have been performed using a realistic coupled channel code~\cite{ha97}.
The results of 
simplified coupled channels code
CCMOD~\cite{da92} which is a modified version of the code 
CCDEF~\cite{da87,ni89} are also presented for comparison.

\subsection{No coupling calculations}

The measured excitation function and the extracted barrier distribution is 
presented in figure~1 along with the predictions of a single--barrier 
penetration model (no coupling). The best fit to the high energy data
requires a diffuseness of 0.85~$fm$, and yields values of
$V_b$~=~74.8~MeV, $R_b$~=~11.6~$fm$ for the barrier height and position 
respectively. As in this case, a fit to the fusion data for other 
systems~\cite{le95}, with calculations in the no--coupling limit also 
required  large values of the diffuseness parameter.
However, it has been shown~\cite{es96} explicitly for 
the $^{16}$O+$^{144}$Sm system  that the large diffuseness
is a result of not considering the couplings to all orders. This seems to be 
true for the $^{16}$O+$^{208}$Pb system too, as 
calculations (i) with $a$~=~0.85~$fm$ 
and no couplings and (ii) with $a$~=~0.65~$fm$ but including the 
2--phonon couplings (see section 3.3), fit the high energy data equally well. 
The best fit Woods--Saxon potential parameters for the latter, obtained by 
fixing $a$~=~0.65~fm are, 
$V_0$~=~235.5~MeV, $R_0$~=~1.1~$fm$ for the 
depth and radius parameter, yielding barrier parameters of 
$V_b$~=~75.2~MeV, $R_b$~=~11.85~$fm$ and $\hbar w$~=~5.0~MeV.  All the 
calculations in this paper have been performed with these parameters.

The calculations which do not include any couplings underpredict the 
excitation function as also 
indicated by the comparison of barrier distributions, where experimentally 
there is significant strength below (and above) the single--barrier.
The failure of the no--coupling calculations to 
reproduce the wide experimental barrier distribution 
clearly indicates that couplings with  other channels need to be considered.

\subsection{Couplings to single phonon states}

Comprehensive coupled channel analysis of elastic, inelastic and fusion 
cross--sections were performed by Thompson \etal~\cite{th89} with the 
inclusion of the lowest 
2$^+$, 3$^-$ and 5$^-$ states of \ipb and 3$^-$ state of \ox in addition to 
neutron pick--up, proton stripping and $\alpha$--transfer channels. 
In this paper, only the inelastic excitations have been included in 
the coupled channel calculations.
It was shown in reference~\cite{th89} that the $\alpha$--transfer channel, 
despite its large 
cross-section,  has very little effect on the fusion cross-section. 
Further, past studies~\cite{mo94,st95,le95,bi96,th89} which have included 
couplings to transfer channels, show that while 
couplings to $Q>0$ transfer channels increase the sub--barrier 
cross--sections, they do not significantly effect the cross--sections at 
higher beam energies; 
the effect of 
$Q<0$ transfer channels is less significant when
couplings to inelastic channels are present. Thus, it is expected that 
for the $^{16}$O~+~$^{208}$Pb system, where the single--nucleon transfers 
Q--values are negative, the dominant features of the fusion barrier 
distribution will be due to couplings to inelastic channels.

The solid lines in figure~1 shows results of realistic coupled channels 
calculation~\cite{ha97} which includes couplings to 
the 2$^+$, 3$^-$ and 5$^-$ vibrational states in \ipb.  
Coupling to 
the 3$^-$ state of \ox has not been considered in any of the calculations 
presented here, as it gives rise to a shift in the barrier distribution 
without changing its shape i.e., in this reaction its effects are only to 
renormalize the real potential. The equivalent CCMOD~\cite{da92} calculations
 are 
shown by the dashed line for comparison; the differences are due to the 
approximations inherent in a simplified coupled channels calculation, mainly
the linear coupling approximation (derivative form--factors) and the 
approximate treatment of 
excitation energies of the intrinsic states.
As seen from the figure, the coupled channel calculations 
fail to reproduce the experimental barrier distribution and the low energy 
part of the 
excitation function. The calculations predict a double peaked structure 
as opposed to the more complex structure seen experimentally; the 
calculations miss the barrier strength 
at the lowest energies and also at around 77~MeV. 
  It should be 
noted that the double--peaked structure of the calculated barrier 
distribution will remain essentially unchanged even when couplings to 
other single--phonon states is considered. The agreement cannot be improved 
by increasing the coupling strength which, while 
decreasing the weight of the lower barrier will simultaneously shift the 
higher barrier to still higher energies. It is thus clear that couplings to 
single phonon states in $^{208}$Pb are not sufficient to explain the data.

\subsection{Couplings to double phonon states} 

As detailed above, calculations with couplings to 
only 1--phonon states are unable to generate a barrier(s) 
which lies at an energy intermediate between the main 
barrier and the higher barrier predicted by these calculations. Thus some
other mechanism has to be considered. Using an eigenchannel approximation
it has been  shown~\cite{ro93} that whereas 
in the case of coupling to a single phonon 
state, the lower and the higher barrier repel each other, the introduction 
of 2--phonon state results in the separation of the lower two barriers being 
smaller,  and the introduction of a third barrier. In the present case, the 
experimental barrier distribution would seem to indicate this scenario.

The existence of two phonon octupole excitations in \ipb was recently 
shown~\cite{ye96} experimentally. Coupled channel calculations 
including the 2$^+$, 3$^-$, 3$^- \otimes$3$^-$ and 5$^-$ vibrational states 
in \ipb and all the resulting cross--coupling terms e.g., 2$^+ \otimes$3$^-$ 
etc, were performed; the double phonon state was treated
in the harmonic limit.
The results are shown by the solid line in Fig.~2; the equivalent CCMOD 
calculations are shown by dashed line. It is clear that the excitation 
function and the shape of the barrier distribution is better reproduced 
in the 2--phonon calculations compared with the 1--phonon calculations.
However, the lower energy part of the excitation function and barrier 
distribution is still not reproduced. A priori, one might assign the   
disagreement to be caused by 
ignoring the couplings to transfer channels, since it is 
known~\cite{mo94,bi96} that coupling to positive Q--value transfer channels 
can introduce a barrier at lower energies. Even though it is recognised that 
yields of transfer cross--sections do not necessarily correlate with 
their coupling to the elastic channel, it is relevant to point out that
for the case of $^{16}$O~+~$^{208}$Pb, the  n--, p-- and 
$\alpha$--transfer reactions observed~\cite{vi77} at energies near the 
barrier have negative Q--values. This could be taken as an 
indication  that the low energy shoulder is unlikely to be due to couplings 
to transfer channels. 
Further, a lower barrier with significant strength and close to the main 
barrier requires a strongly coupled channel and since 
the coupling between the elastic and transfer channels 
are generally small in comparison with inelastic channels, it would be 
difficult to reproduce the observed barrier distribution by couplings to 
transfer channels only. Due to limitations of the coupled channels code, at 
present we are unable to perform the calculations including transfer channels
to investigate these suggestions.

\section{Conclusions}

Comparison of the experimentally measured barrier distribution with the
results of coupled channel calculations show that coupling to 
double octupole phonon excitations in \ipb are necessary to explain the 
fusion cross--sections. 
The 
experiment shows the presence of another lower barrier which could not 
be reproduced by these  calculations. While this barrier(s) might arise 
due to couplings to transfer channels, not included in the present 
treatment, it is unlikely to reproduce the shape of the barrier 
distribution as discussed in section 3.3. One might have to look for 
other reasons like effects of 
anharmonicities of the 2--phonon state in \ipb as discussed by Takigawa  
\etal during this conference. 

It is interesting that the dynamics of the fusion process even for reactions 
between two closed shell nuclei, which might be thought to be a simple 
process particularly at low energies, is affected by complex surface 
vibrations like the 2--phonon states. Furthermore, the barrier distribution 
picture indicates that there are other subtle features of the 
$^{16}$O + $^{208}$Pb reaction which have yet to be 
understood, indicating that experiments are being done at a level of 
precision where a better understanding of the approximations in the 
theoretical calculations is required.

\newpage

\begin{figure}

\caption{ Measured and calculated fusion excitation functions and 
barrier 
distributions for the \ox + \ipb system. The curves are the result of using a
coupled channels code with no coupling (dotted), and couplings to 2$^+$, 3$^-$ 
and 5$^-$ states of \ipb (solid line). Calculations using the simplified 
coupled channels code CCMOD for the same couplings  is shown by the dashed 
line.}

\caption{ Comparison of the experimental data with 
calculations including couplings up to two phonon states (see section 3.3) 
in \ipb. Calculations with both the exact coupled channel code (solid line)
and the simplified coupled channel code CCMOD (dashed line) are shown.}  

\end{figure}


\section*{References}

\begin{thebibliography}{fred}
\bibitem{ro91}Rowley N, Satchler G R and
Stelson P H 1991 {\it Phys. Lett.} {\bf B254} 25 

\bibitem{da83}Dasso C H, Landowne S and Winther A 1982
{\it Nucl. Phys. A}   {\bf 405} 381 (1982); 1983 {\bf 407} 221

\bibitem{xi91}Wei J X, Leigh J R, Hinde D J, Newton J O, 
Lemmon R C, Elfstr\"{o}m S, Chen J X and Rowley N 1991 {\it Phys. Rev. 
Lett.} {\bf 67} 3368 

\bibitem{mo94}Morton C R, Dasgupta M, Hinde D J, Leigh J R, Lemmon R C,
Lestone J P, Mein J C, Newton J O, Timmers H, 
Rowley N and Kruppa A T 1994
{\it Phys. Rev. Lett.} {\bf 72} 4074 


\bibitem{st95}Stefanini A M, Ackermann D, Corradi L, Napoli D R, 
Petrache C, Spolaore P, Bednarczyk P, Zhang H Q, Beghini S, Montagnoli G,
Mueller L, Scarlassara F, Segato G F, Soramel F and Rowley N,
1995 {\it Phys. Rev. Lett.} {\bf 74} 864 



\bibitem{le95}Leigh J R, Dasgupta M, Hinde D J,
Mein J C, Morton C R, Lemmon R C, Lestone J P, Newton J O, Timmers H,
Wei J X and Rowley N 1995 {\it Phys. Rev. C} {\bf 52} 3151 


\bibitem{bi96}Bierman J D, Chang P, Liang J F, Kelly M P, 
Sonzogni A A and Vandenbosch R 1996 {\it Phys. Rev. Lett.}
{\bf 76} 1587 

\bibitem{th89}Thompson I J, Nagarajan M A, Lilley J S and Smithson M J
1989 {\it Nucl. Phys. A} {\bf 505} 84

\bibitem{mo95}Morton C R, Hinde D J, Leigh J R,
Lestone J P, Dasgupta M, Mein J C, Newton J O and
Timmers H 1995 {\it Phys. Rev. C} {\bf 52} 243

\bibitem{ha97}Hagino K, Takigawa N, Dasgupta M, Hinde D J, 
Leigh J R 1997 {\it Phys. Rev. C} {\bf 55} 276

\bibitem{da92}Dasgupta M, Navin A,
Agarwal Y K, Baba C V K, Jain H C,
Jhingan M L and Roy A 1992 {\it Nucl. Phys. A} {\bf 539} 351



\bibitem{da87}Dasso C H and Landowne S 1987 {\it Comp. Phys. Commun.}
 {\bf 46} 187 

\bibitem{ni89}Fern\'{a}ndez Niello J O, Dasso C H and Landowne S 1989, 
{\it Comp. Phys. Commun.} {\bf 54} 409 

\bibitem{es96}Esbensen H and Back B B 1996 {\it Phys. Rev. C} {\bf 54} 3109



\bibitem{ro93}Rowley N and Dasgupta M 1993 {\it Proc. 
Int. Conf. on Heavy--Ion Reactions with Radioactive Beams (RIKEN)}
(Singapore: World Scientific Press) p~232


\bibitem{ye96}Yeh Minfang, Garrett P E, McGrath C A, Yates S W and
Belgya T 1996 {\it Phys. Rev. Lett.} {\bf 76} 1208

\bibitem{vi77}Videb\ae k F, Goldstein R B, Grodzins L and  Steadman S G 
1977 {\it Phys. Rev. C} {\bf 15} 954


\end{thebibliography}
\end{document}